# Optical Deformations in Solar Glass Filters for High Precision Astrometry.

Costantino Sigismondi, ICRA and G. Ferraris Institute, Rome, Italy.
sigismondi@icra.it
Alexandre H. Andrei ON OV/UFRJ, Rio de Janeiro, Brazil.
Sérgio Boscardin ON, Rio de Janeiro, Brazil.
Jucira L. Penna ON, Rio de Janeiro, Brazil, and
Eugênio Reis-Neto MAST, Rio de Janeiro, Brazil.

Measuring the solar diameter at all position angles gives the complete figure of the Sun. Their asphericities have implications in classical physics and general relativity, and the behavior of the optical systems used in the direct measurements is to be known accurately.

A solar filter is a plane-parallel glass with given absorption, and here we study the departures from the parallelism of the faces of a crystal slab 5 mm thick, because of static deformations. These deformations are rescaled to the filter's dimensions.

Related to the Solar Disk Sextant experiment and to the Reflecting Heliometer of Rio de Janeiro a simplified model of the influences of the inclination between the external and the internal surfaces of a glass solar filter, is discussed.

Keywords Astrometry, Solar Diameter, Optical distortions

**Introduction: the figure of the Sun in General Relativity and classical physics**

A long debate about the sphericity of the Sun arose to test alternative theories of General Relativity. The anomalous precession of the perihelion of Mercury, discovered in the second half on nineteenth century by Urbain Joseph Le Verrier in Paris and furtherly studied by Simon Newcomb in Washington DC at the Naval Observatory, was fully explained only by Albert Einstein in the Theory of General Relativity.

The effect of the curved space by the gravity of the Sun upon the planets could be either the result of the attraction of an inner planet, which Le Verrier named Vulcanus, but was never found, either the effect of a dust ring around the Sun or finally an oblateness of the Sun itself.

This last hypotheses remained after the dismal quest of Vulcanus and dust belts around the Sun, and gave origin to the researches of Robert Dicke in Princeton during the years 60s and 70s of the twentieth century.

For a more detailed analysis of these historical processes see Sigismondi (2011).

**Historical instruments: Göttingen Heliometer and SDS experiment (1984-2011)**

The design of the Heliometer of Göttingen changed the original scheme of Dollond and perfectioned by Joseph Fraunhofer: instead of cutting a lens into two parts and shifting them using a micrometer (R. C. Brooks, 1991) Schur and Ambronn (1895) realized a glass wedge to be put in front of the objective of the telescope, realizing two images of the Sun, the direct and the first reflected one, whose separation is related to the solar diameter by the constancy of the wedge's angle.

The same principle has been repeated in the experiment of Solar Disk Sextant, designed in 1984 and launched from 1992 to 2011 in 6 useful flights to the top of the stratosphere.

The stability of the wedge's angle along the years allowed to compare the measurements of the solar diameter in the various flights.

This design helps to understand the effect of converging or diverging surfaces of the glass filters to be used in solar astrometry: each part of the filter acts like a wedge with a tiny angle and producting a secondary image superimposed with the direct one, which changes slightly the position of the inflection point, used in the FFT algorithm to define the solar limb.

**The Reflecting Heliometer of Rio de Janeiro**

The concept of this new Heliometer, conceived by Victor d'Avila and realized in Rio de Janeiro at the Observatorio Nacional (ON) by the Solar Physics group cohautors of this work, is based upon the neutrality of the reflection's law with respect to the wavelengths.

A parabolic mirror is split into two parts along a diameter and the two parts are inclined of a given angle along the direction perpendicular to the first one.

The images produced by each part of the mirror behave like the two heliometric classical images and their distance is linked to the actual solar diameter.

The possibility to rotate the Heliometer along the optical axis of the original mirror allows to measure the solar diameter at all position angles, to recover the figure of the Sun.

A new version of this Reflecting Heliometer is the annular one, where from the initial mirror is cut a central circular mirror and the left corona is rotated along a diameter of the splitting angle required to have two heliometric images to recover the solar diameter at a given position angle on the Sun.

This second version is under construction at the Observatório Nacional in Rio de Janeiro.

**The asphericities of the Sun after RHESSI[1] and Picard satellites**

The results of RHESSI in 2005 and PICAR in 2014 have confirmed the previous measurements on the figure of the Sun started with SDS experiments and other measurements made on the ground, like the solar astrolabes, the utmost accurate experiments conducted on the ground on the measurements of the solar diameter.

The evolution of the parameters like diameter and oblateness is still controversial either because of its astrophysical consequences either because the errorbars of all experiments are still enough big to exclude any variation at all.

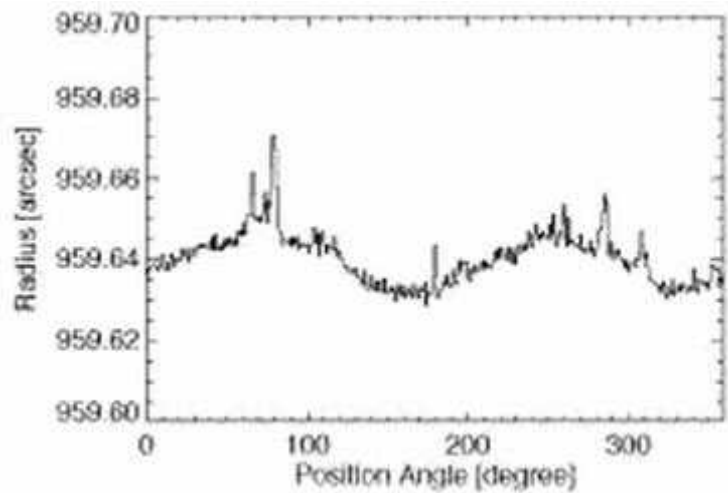

The results of RHESSI experiment (Fivian et al. 2005) are summarized in figure 1, the equatorial excess is 6.8 milliarcsec; the results of PICARD (Irbah et al. 2014) are 8.4 milliarcsec of oblateness, i.e. the polar diameter is 6.1 km smaller than the equatorial one, due to the solar rotation.

**Departures from plane parallel geometry of a crystal slab**

This experiment wants to quantify the departures from plane parallel geometry of a crystal slab, in order to estimate the same features on a glass filter.

The slab is 89 cm x 52.5 cm, 0.45 cm thick and 6 Kg, made by bohemian crystal.

It is located on a wooden table, since approximately 100 years.

The density is 2.8 g/cm³.

A LASER beam of λ=625 nm is used to produce reflected images: the first from the upeer surface, the second one from the lower one. Higher level reflections are neglected.

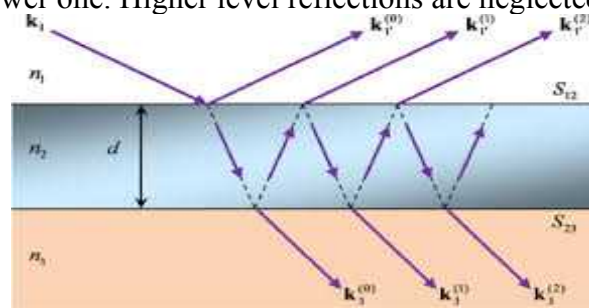

---
1   http://science.nasa.gov/missions/rhessi/

On the ceiling of the room where the experiment has been realized the distance between the two spots of the LASER beam have been measured; the measurements have been made along both the long sides of the slab and on the central axis, and along the short sides of the slab and on the central axis perpendicular to the previous one.

Where the two images are gathering into one the lower surface is bent toward the upper one, while when the two dots are separating each other the lower surfaces separates from the upper one in the direction of the beams.

The results are the following: on the short right R side moving from the beginning to the end and projecting the LASER beam from 50 cm, at the same height above the slab a distance of 1 cm perpendicular to the direction of the beam (parallel to the short sides of the slab) is measured.

In the center of the slab the two dots are almost superimposed.

At the left edge L the distances of the two spot starts from 0.5 cm, perpedicular to the direction of the beam, arriving at the end almost superimposed.

The central part is rather parallel with respect to the accuracy of our measurements, the R part is divergent from the beginning to the end with a constant angle (always with respect to the accuracy of our measurements) and the L part is divergent from right to left end at the beginning and recovers the parallelism on the end.

To infer the statical deformations to which the slab has been subjected being AB the long side (beginning point where the light is cast) and CD the long side (ending points) the corners D and B are at the same level while A and C could be at a slightly lower levels.

The split into two dots at 1 cm distant after 50 cm implies an angle of divergence of about 1/100 rad i.e. 30'. A split of 0.5 cm a divergence of 15'.

**Rescaling to filter's dimensions**

Over a slab of 1 m x 0.5 m small differences in the level of the four corners produces deformations of 15' to 30' where the lower wooden level decreases its sustain on the glass.

Possibily the lower surface of the slab is more inclined of the upper one.

A glass solar filter 8", say a circle of 0.2 m of diameter, is mounted on aluminum frame and has a total weight of 0.5 Kg, of which about 200 g of glass.

The stresses due to the weigth are smaller, but the difference in quote of the glass and in compression by the frame are much larger than our slab of crystal.

The thickness of the glass filter is 2.3 mm, undergoing gravitational stresses 1/30 of the big slab in the presence of anisothropies in the aluminum frame (different pressures on the glass or different quotes).

If the deformation is linear with the time, after 100 years a deformation of 1/30 of 30' can be expected, which is 1', or 60". And in the first years of use of the filter 0.6" per year of deformation can be reasonably expected.

**Conclusions: the influences of departures from parallelism**

The departures from the parallelism in the solar filter act like the wedge for the SDS experiment.

There is the production of a secondary image, fainter than the first one, partially superimposed to the primary one, which presence affects the identification of the inflection point position.

If the bending is constant along the whole glass filter this effect is the rising of the measured solar diameter exactly by the angle given by the inclination of the surfaces, but in the case more realistic of variable inclination along the whole disk a combination of duplicate images with different angle has to be computed and weighted with the correspondant area where a given inclination prevails.

Just for the sake of clarity the following examples: if there are two halves of the filter with opposite inclinations the net result is a broadening of the slope at the inflection point from either sides, but when the filter is rotated of 90° the effect disappears.

A complex deconvolution should be taken into account if the solar diamters changes with rotation of the axis of the instrument.